\documentclass[aps,preprint]{revtex4}%
\usepackage{amsfonts}
\usepackage{amsmath}
\usepackage{amssymb}
\usepackage{graphicx}%
\begin{document}
\title{\textbf{The temperature in Hawking radiation as tunneling }}
\author{Baocheng Zhang$^{a,b}$}
\author{Qing-yu Cai$^{a}$}
\author{Ming-sheng Zhan$^{a}$}
\affiliation{$^{a}$State Key Laboratory of Magnetic Resonances and Atomic and Molecular
Physics, Wuhan Institute of Physics and Mathematics, The Chinese Academy of
Sciences, Wuhan 430071, People's Republic of China}
\affiliation{$^{b}$Graduation University of Chinese Academy of Sciences, Beijing 100081,
Peope's Republic of China}
\author{}

\begin{abstract}
The quasi-classical method of deriving Hawking radiation under the
consideration of canonical invariance is investigated. We find that the
horizon should be regarded as a two-way barrier and the ingoing amplitude
should be calculated according to the negative energy particles tunneling into
the black hole because of the whole space-time interchange and thus the
standard Hawking temperature is recovered. We also discuss the advantage of
the Painlev\'{e} coordinates in Hawking radiation as tunneling.

PACS classification codes: 04.70.Dy, 04.20.Fy

Keywords: Horizon; Tunneling; Time coordinate; Radiation temperature

\end{abstract}
\maketitle

\section{\textbf{Introduction}}

A classical black hole can only absorb and not emit particles. When
considering quantum effect, however, Hawking discovered \cite{swh75} that
black hole emits thermal radiation with a temperature $T=\frac{\kappa}{2\pi}$,
where $\kappa$ is the surface gravity of black hole. The physical reason of
radiation was explained \cite{hh76} as coming from vacuum fluctuations
tunneling through the horizon of the black hole. But some original derivation
based on the Bogoliubov transformation \cite{swh75} or other methods
\cite{hh76,dnp76,wgu76} didn't have the direct connection with the view of
tunneling. Moreover, these methods, in which the background geometry is
considered fixed, didn't enforce the energy conservation during the radiation
process. Recently Parikh and Wilczek suggested \cite{pw00} a method based on
energy conservation by calculating the particle flux in Painlev\'{e}
coordinates from the tunneling picture. Their result recovered the Hawking's
original result in leading order and gave the consistent temperature
expression and the entropy relation. The method had also been discussed
generally in different situations \cite{rzg05,amv05,jwc06,ma06,sk08} and
showed the formula was self consistent even checked by using thermodynamic
relation \cite{bm081,bm082,bms08}. Another importance is to give the
non-thermal spectrum which implies there may exist the information-carrying
correlation in the radiation.

Another method called Hamilton-Jacobi method \cite{sp99} had also been
proposed to obtain the tunneling probability besides the radial null geodesic
method \cite{pw00}. But by using the Hamilton-Jacobi equation, the Ref.
\cite{aas06} gave the temperature twice as large as the Hawking temperature.
Although the double temperature didn't affect the connection of black hole
radiation with the thermodynamic law as long as the proportional relation
between the temperature and the surface gravity is held \cite{zcz08}, it will
also cause the change of entropy and radiation temperature which is observable
by astrophysical method or at LHC. So determining whether the temperature is
twice or whether there exists the factor of 2 problem is important. There had
been two methods suggested to solve this problem and concluded the temperature
was the same as the Hawking temperature. But the two methods look as if they
were different completely. One of them \cite{pm07} explained that the standard
temperature could be obtained by using the detailed balancing formula that is
the ratio of the outgoing and incoming probabilities, in which the canonical
invariance or the tunneling dependent on direction of move was not considered.
The other one \cite{apgs08,aps08} pointed out that one must take into account
the ignored temporal contribution of the action in order to recover the
original Hawking temperature, in which the normalization of the ingoing
probability was not involved and the temporal rotation at the Schwarzschild
horizon is periodic and indefinite. So it is necessary to find another method
which not only overcomes these deficiencies but also recover the standard
Hawking temperature.

In our paper we discuss the property of the horizon and regard the horizon as
a two way barrier when one considers virtual particle pairs inside and outside
the horizon. According to the propagator theory, we find that when one treats
the black hole radiation as tunneling, the ingoing amplitude should be
calculated as tunneling of negative energy particles. Thus we can recover the
standard Hawking temperature by using the canonically invariant tunneling
transmission rate, $\Gamma=e^{-\operatorname{Im}\left(
{\displaystyle\oint}
pdx\right)  }$. We show that the Painlev\'{e} coordinates is more convenient
than Schwarzschild coordinates for calculating the Hawking temperature in the
picture of tunneling. In the end we also discuss the other two methods
\cite{pm07,apgs08,aps08} and compare them with our method.

In this paper we take the unit convention $k=\hbar=c=G=1$.

\section{Tunneling, horizon and propagator}

Tunneling is a quantum mechanical phenomenon to happen when the initial and
final states are separated by a barrier which cannot be classically crossed
because the system does not have enough energy. Generally speaking, there are
two kinds of tunneling which can be described as one when the barrier is
insensitive to the direction of motion and the other one when the barrier is
sensitive to the direction of motion. For the former there are two equivalent
expressions for the tunneling transmission coefficient, $\Gamma
=e^{-2\operatorname{Im}\left(  \int pdx\right)  }=e^{-\operatorname{Im}\left(
%
{\displaystyle\oint}
pdx\right)  }$; for the latter only one of the expressions is applicable under
invariance of canonical transformation, $\Gamma=e^{-\operatorname{Im}\left(
{\displaystyle\oint}
pdx\right)  }$. Generally we do not have this problem in the usual examples of
tunneling for the form, $\Gamma=e^{-2\operatorname{Im}\left(  \int pdx\right)
},$ which is because for those situations the tunneling in both direction is
equally suppressed \cite{bdc08}. But for other situations such as the
tunneling through black hole horizon, we have to notice this problem because
the tunneling through black hole horizon is sensitive to the direction of
motion. The classical infalling particles face no barrier at all and cross the
horizon freely but the classical outgoing particles are forbidden or cannot
cross the horizon. When considering the quantum effect, however, the vacuum
fluctuation can lead to generate the virtual pairs of negative-positive energy
particles and makes the tunneling possible. The tunneling includes two parts,
one of which is the positive energy particles tunnel out of the horizon for
the virtual pair inside the horizon and the other one of which is the negative
energy particles tunnel inward through the horizon for the virtual pair
outside the horizon. Thus the horizon represents a two way barrier when one
considers virtual particle pairs inside and outside the horizon. So we should
choose $\Gamma=e^{-\operatorname{Im}\left(
{\displaystyle\oint}
pdx\right)  }$ as proper observable for the tunneling through black hole
horizon \cite{aas06}.

On the other hand, the propagator as a transition amplitude \cite{jjs05} is
symmetric under interchange of space-time coordinates. The propagator can be
written as $K(x_{f}t_{f};x_{i}t_{i})=\left\langle x_{f},t_{f}|x_{i}%
,t_{i}\right\rangle =e^{-iS}$, where the quantum system is transferred from
the initial place and time $x_{i},t_{i}$ to the final place and time
$x_{f},t_{f}$, $t_{f}>t_{i}$ and $S$ is the action of the system. So we can
gain the propagator under interchange of space-time coordinates as%
\begin{equation}
K(x_{i}t_{i};x_{f}t_{f})=\left\langle x_{i},t_{i}|x_{f},t_{f}\right\rangle
=e^{iS^{+}}=e^{i\operatorname{Re}S}e^{\operatorname{Im}S},
\end{equation}
and%
\begin{equation}
K(x_{f}t_{f};x_{i}t_{i})=\left\langle x_{f},t_{f}|x_{i},t_{i}\right\rangle
=e^{iS}=e^{-i\operatorname{Re}S}e^{\operatorname{Im}S}.
\end{equation}
This shows that the propagator is equivalent under interchange of space-time
coordinates up to a pure imaginary phase. When the Hamiltonian is independent
on the time, the action is separable for the time and space coordinates,
$S=Et+S_{0}(x)$. And when the system can be treated reliably in the
short-wavelength limit, the WKB approximation can be used. For the Hawking
radiation as tunneling, the conditions of WKB approximation are satisfied
because the Schwarzschild space-time is stationary and the particles or the
short wavelength limit is supported due to the infinite blueshift of the
outgoing wave-packet near the horizon \cite{pw04}. The propagator can be also
be written as $K(x_{f}t_{f};x_{i}t_{i})=\left\langle x_{f}\right\vert
\exp(-iH(t_{f}-t_{i}))\left\vert x_{i}\right\rangle $, where $H$ is the
Hamiltonian of the particles tunneling outward in black hole radiation. Since
the particle with energy $E$ is considered, we have%
\begin{equation}
K(x_{f}t_{f};x_{i}t_{i})=\left\langle x_{f}\right\vert e^{-iH(t_{f}-t_{i}%
)}\left\vert x_{i}\right\rangle =\left\langle x_{f}|x_{i}\right\rangle
e^{-iE\Delta t},
\end{equation}
and%
\begin{equation}
K(x_{i}t_{i};x_{f}t_{f})=\left\langle x_{i}\right\vert e^{-iH(t_{i}-t_{f}%
)}\left\vert x_{f}\right\rangle =\left\langle x_{i}|x_{f}\right\rangle
e^{iE\Delta t}=\left\langle x_{i}|x_{f}\right\rangle e^{-i(-E)\Delta t},
\end{equation}
where the amplitude $\left\langle x_{f}|x_{i}\right\rangle $ and $\left\langle
x_{i}|x_{f}\right\rangle $ can be calculated in the semiclassical
approximation and $\Delta t=t_{f}-t_{i}>0$. The amplitude can be used to
describe the tunneling probability, $\Gamma=\left\langle x_{f}|x_{i}%
\right\rangle \left\langle x_{i}|x_{f}\right\rangle =\exp\left(
\operatorname{Im}\left(
{\displaystyle\oint}
pdx\right)  \right)  $. However, when $\left\langle x_{f}|x_{i}\right\rangle $
and $\left\langle x_{i}|x_{f}\right\rangle $ are regarded as tunneling
amplitude and their time is going on according to $t_{i}\rightarrow t_{f}$, it
is noted that for ingoing amplitude $\left\langle x_{i}|x_{f}\right\rangle $
the energy of the tunneling particles must be treated as negative. Actually
the ingoing amplitude is obtained along the reversed time, so according to
Feyman's idea that negative-energy particles can only travel backward in time,
the energy of the tunneling particles for ingoing amplitude should also be
negative. Therefore, when we calculate the closed contour integral, the
outgoing amplitude and ingoing amplitude have to be calculated as tunneling of
particles with the opposite energy. Noticed that the horizon as barrier is
single-directional for particles with the same energy. Here the horizon as
barrier can be treated as both-directional for particles with the opposite
energy due to the consideration of the temporal interchange of outgoing and
ingoing amplitude, but this doesn't mean the both direction is equally
suppressed. We have to calculate them respectively. Along this line we will
recover the temperature of black hole radiation in the next section.

\section{The temperature}

For a particle, of mass, $m$, the Hamilton-Jacobi equation is%
\begin{equation}
g^{\mu\nu}\partial_{\mu}\partial_{\nu}S+m^{2}=0, \label{hje}%
\end{equation}
where $g^{\mu\nu}$ is the inverse metric of the background space-time and $S$
is the action of the particle. Thus one can express the scalar field as
$\phi(x)=\exp[-\frac{i}{\hbar}S+\cdots]$. In the picture of Hawking radiation
as tunneling, the Painlev\'{e} coordinates is considered as appropriate
because it, unlike Schwarzschild coordinates, is not singular at the horizon.
The barrier is created by the outgoing particles themselves, which is ensured
by the energy conservation \cite{pw04}. We can express the Painlev\'{e}
coordinates as%
\begin{equation}
ds^{2}=-(1-\frac{2M}{r})dt_{p}^{2}+2\sqrt{\frac{2M}{r}}drdt_{p}+dr^{2}%
+r^{2}d\Omega^{2}. \label{pm}%
\end{equation}

Since the metric is stationary and has a time-like Killing vectors, we can
split the action into a time and spatial part, $S=Et_{p}+S_{0}(r),$ where $E$
is the energy of particle. We use Eq. (\ref{hje}) and obtain%
\begin{equation}
S_{0}(r)=-\int\frac{dr}{1-\frac{2M}{r}}\sqrt{\frac{2M}{r}}E\pm\int\frac
{dr}{1-\frac{2M}{r}}\sqrt{E^{2}-m^{2}(1-\frac{2M}{r})}, \label{pa}%
\end{equation}
where the positive and negative sign indicates that the particle is ingoing
and outgoing. Note that the contour integral includes a singularity at $r=2M$
and it has to be made by going around the pole at singularity. In Ref.
\cite{aas06}, the result is obtain as $\operatorname{Im}S_{0}(r)=0$ for the
ingoing particles (which corresponds to the plus sign in Eq. (\ref{pa})) since
the first and second terms have the same magnitude and $\operatorname{Im}%
S_{0}(r)=-4\pi ME$ for the outgoing particles and the authors conclude that
the temperature $T=\frac{1}{4\pi M}$ can be obtained by comparing the
tunneling probability $\Gamma=\exp(\operatorname{Im}\left(
{\displaystyle\oint}
pdx\right)  )=e^{-4\pi ME}$ with a Boltzmann factor $\Gamma=\exp(-\frac{E}%
{T})$. The temperature is twice as large as the original Hawking temperature
and it is the same for the other coordinates (Schwarzschild, isotropic and so
on), which seems to imply that one should discard the Hawking's original
calculation. It is not the case, however. We note that the calculation above
for the ingoing particles is concerned about the positive energy particles and
according to our analysis in the last section when considering the temporal
interchange the energy of ingoing particles should be treated as negative. So
$\operatorname{Im}S_{0}(r)=4\pi ME$ for the ingoing particles, which is
consistent with that of the negative energy particles tunneling inward
calculated in Ref. \cite{pw00}. It should be stressed that here $E$ is always
larger than zero and when we consider the negative energy particle, the minus
sign before the first term in Eq. (\ref{pa}) has to be changed to plus sign
and so the calculation becomes the addition of the two equivalent terms. Thus
we can obtain the tunneling probability as%
\begin{equation}
\Gamma=e^{\operatorname{Im}\left(
{\displaystyle\oint}
pdr\right)  }=e^{\operatorname{Im}\left(  \int p_{r}^{out}dr-\int p_{r}%
^{in}dr\right)  }=e^{-8\pi ME}.
\end{equation}
And in the same way we associate it with a Boltzmann factor $\Gamma
=\exp(-\frac{E}{T})$, so the temperature $T=\frac{1}{8\pi M}$, which is the
standard temperature obtained by Hawking.

If the Schwarzschild metric is used, this yields $\operatorname{Im}%
S_{0}(r)=\pm2\pi ME$ \cite{aas06}. But we observed that when the plus sign is
taken, the ingoing amplitude is not a decay but is an amplification. Thus in
the classical limit ($\hbar\rightarrow0$) the tunneling will not disappear and
trends to infinity \cite{pm07}, which is inconsistent with our experiential
fact that the tunneling is a kind of quantum effect and doesn't occur in the
classical field. Therefore one must use the integral constant to adjust the
amplitude and for the positive energy particle tunneling outward through the
horizon, we have%
\begin{equation}
\operatorname{Im}S_{pe}^{out}=-2\pi ME+C,
\end{equation}%
\begin{equation}
\operatorname{Im}S_{pe}^{in}=2\pi ME+C,
\end{equation}
where the label $pe$ means the tunneling energy is positive and $C$ is a
constant. In order to avoid the infinity problem in the classical limit and
ensure that the amplitude is unity in the classical limit where everything is
absorbed, we have to take $C=-2\pi ME$. That is to say that the amplitude
$\left\langle in|out\right\rangle _{pe}=\exp(i\theta)$, where $\theta$ is an
arbitrary phase and may be related to the horizon when there existed the
quantum fluctuation. Thus $\operatorname{Im}S_{pe}^{out}=-4\pi ME$ and so the
amplitude is gotten as%
\begin{equation}
\left\langle out|in\right\rangle _{pe}=\exp(-i\theta)\exp(-4\pi ME).
\end{equation}
The constant $C$ occurred may be due to the \textquotedblleft
badness\textquotedblright\ of the Schwarzschild coordinates near the horizon.
It is noted if we choose the Painlev\'{e} coordinates to calculate, the result
is $\operatorname{Im}S_{pe}^{out}=-4\pi ME$ and $\operatorname{Im}S_{pe}%
^{in}=0$ and so the normalization is not needed. This implies that the
Painlev\'{e} coordinates indeed is a good choice which not only behaves well
at the horizon but also is convenient to obtain the temperature even by
Hamilton-Jacobi method.

Similarly for the negative energy particle tunneling inward through the
horizon, we have%
\begin{equation}
\operatorname{Im}S_{ne}^{out}=2\pi ME+D,
\end{equation}%
\begin{equation}
\operatorname{Im}S_{ne}^{in}=-2\pi ME+D,
\end{equation}
where the label $ne$ means the tunneling energy is negative and $D$ is a
constant. The constant introduced is the same reason as that above and if we
choose the Painlev\'{e} coordinates, the constant is not necessary. Here
suppose that $E$ is positive and so for the negative energy we have to replace
$E$ by $-E.$ The same reason makes us take $D=-2\pi ME.$ It shows that in
classical limit the negative energy particles can only move out of the black
hole and so the mass or area of black hole never decrease which is consistent
with the second law of classical black hole thermodynamics \cite{bch73}.
Therefore the amplitude $\left\langle out|in\right\rangle _{ne}=\exp
(-i\theta)$. On the other hand, $\operatorname{Im}S_{ne}^{in}=-4\pi ME$ and so
the ingoing amplitude is gotten as%
\begin{equation}
\left\langle in|out\right\rangle _{ne}=\exp(i\theta)\exp(-4\pi ME).
\label{net}%
\end{equation}
From the analysis above one knows that $\left\langle in|out\right\rangle
^{\ast}\neq\left\langle out|in\right\rangle $ if we only consider the positive
or negative energy particles tunneling. However, it is noted that the
calculation of $S^{in}$ and $S^{out}$ only includes the space coordinates
change, that is to say the integral is made from $r_{out}\rightarrow r_{in}$
and $r_{in}\rightarrow r_{out}$, but the time coordinates change is not
considered. According to our analysis in the last section, when calculating
the tunneling probability by the closed contour integral we have to take the
amplitude $\left\langle out|in\right\rangle _{pe}$ and $\left\langle
in|out\right\rangle _{ne}$ and so we gain%
\begin{equation}
\Gamma=\left\langle out|in\right\rangle _{pe}\left\langle in|out\right\rangle
_{ne}=\exp(-8\pi ME).
\end{equation}
And the temperature is $T=\frac{1}{8\pi M}$ which is also consistent with the
Hawking's original result. On the other hand, it is noted that when
considering the problem of tunneling inward we can obtain the probability
$\Gamma=$ $\left\langle in|out\right\rangle _{pe}\left\langle
out|in\right\rangle _{ne}=1$, which is consistent with the fact the ingoing
particles face no barrier.

\section{Discussion}

We also notice that there are two methods suggested to solve this problem.

One suggestion \cite{pm07} is made along the line of path integral in the
complex time analysis \cite{hh76} in which the amplitudes for particle
emission is related to that for particle absorption with the result that the
ratio of emission and absorption probabilities for energy $E$ is given by%
\begin{equation}
P_{emission}=\exp(-\frac{E}{T_{H}})P_{absorption}.
\end{equation}
This formula is used to obtain the Hawking radiation in a new path integral
method and at the same time it also gives the same temperature as Hawking's
original result. In Ref. \cite{pm07}, it is applied to solve the factor of 2
problem about black hole temperature and the problem of the absorption
probability which tends to be greater than unity and goes to infinity in the
classical limit has been pointed out. After normalization, one can find that
$\left\langle in|out\right\rangle _{pe}=1$ and $\left\langle
out|in\right\rangle _{pe}=-4\pi ME$. It is noticed that the author obtains the
emission and the absorption probability directly from the modulus square of
the amplitude while doesn't consider that the tunneling is sensitive to the
direction of motion. According to our suggestion, the emission probability
should be calculated as%
\begin{equation}
P_{emission}=\left\langle out|in\right\rangle _{pe}\left\langle
in|out\right\rangle _{ne}=-8\pi ME,
\end{equation}
and the absorption probability is%
\begin{equation}
P_{absorption}=\left\langle in|out\right\rangle _{pe}\left\langle
out|in\right\rangle _{ne}=1.
\end{equation}
Thus the tunneling probability is gotten as
\begin{equation}
\Gamma=\frac{P_{emission}}{P_{absorption}}=\exp(-8\pi ME),
\end{equation}
and the temperature is $T=\frac{1}{8\pi M}$. It seems that such treatment
gives the same result as that in Ref. \cite{pm07}. But this doesn't mean that
the tunneling is not dependent on the direction of move. Especially whether
the amplitude $\left\langle in|out\right\rangle _{ne}$ is equal to
$\left\langle out|in\right\rangle _{pe}$ for all situations has to be proven
further, but here they are the same.

The second suggestion \cite{apgs08,aps08} is that not only the spatial part
but also the temporal part contributes to the imaginary part of action. In
Schwarzschild background, the spatial contribution to the action is $%
{\displaystyle\oint}
pdr=-4\pi ME$ and the temporal contribution to the action is seen by
transfering the Schwarzschild coordinates into Kruskal-Szekeres coordinates.
The transformation is given as%
\begin{equation}
T=(\frac{r}{2M}-1)^{1/2}e^{r/4M}\sinh(\frac{t}{4M}),R=(\frac{r}{2M}%
-1)^{1/2}e^{r/4M}\cosh(\frac{t}{4M}), \label{skt1}%
\end{equation}
for the region exterior to the black hole $(r>2M)$ and%
\begin{equation}
T=(1-\frac{r}{2M})^{1/2}e^{r/4M}\cosh(\frac{t}{4M}),R=(1-\frac{r}{2M}%
)^{1/2}e^{r/4M}\sinh(\frac{t}{4M}), \label{skt2}%
\end{equation}
for the interior of the black hole $(r<2M)$. To connect these two patches
across the horizon at $r=2M$ one needs to \textquotedblleft
rotate\textquotedblright\ the Schwarzschild $t$ as $t\rightarrow t-2i\pi M$
(together with the change $r-2M\rightarrow2M-r$). So the temporal contribution
is $\operatorname{Im}(E\Delta t^{in,out})=\pm2\pi ME$. By adding the temporal
and spatial contribution, the Hawking temperature is recovered as $T=\frac
{1}{8\pi M}$. This is indeed an ingenious solution. However there are still
some subtle places to be noticed. At first, the time transformation is
periodic (the period is $8i\pi M$) and no necessary reason demands the
\textquotedblleft rotation\textquotedblright\ is $2i\pi M$ (maybe it is $6i\pi
M$, but this is indefinite). In other words, one can also suppose that for the
outgoing amplitude the rotation is $2i\pi M$ and for the ingoing amplitude the
rotation is $6i\pi M$ because they exist within one period, so the total
rotation is $8i\pi M$. In fact, the time axial can be extended from real axial
to virtual axial in Kruskal-Szekeres coordinates, $t=-i\tau$. Thus in the the
region exterior to the black hole the virtual $\tau$ is a coordinate with
period $8\pi M$ and such character satisfies \textquotedblleft thermal Green
function\textquotedblright, $G_{T}(x,t;x_{0},t_{0})\sim G_{T}(x,t+i\beta
;x_{0},t_{0})$ where $\beta=$ $8\pi M=\frac{1}{T}$. We can relate the path
integral propagator with thermal propagator and thus to the observer in static
frame it will seems as if he is in a bath of blackbody radiation at the above
temperature \cite{gp76,gh77}. Secondly, the rotation described in Ref.
\cite{aps08} may be only applicable to the Schwarzschild coordinates, for
other coordinates this has to treated carefully. For example, for the
Schwarzschild and Painlev\'{e} coordinates, $\int Edt+%
{\displaystyle\oint}
pdr=\int Edt_{p}+%
{\displaystyle\oint}
p_{p}dr$ where $%
{\displaystyle\oint}
pdr=%
{\displaystyle\oint}
p_{p}dr$ because of canonical invariance. So $\int Edt=$ $\int Edt_{p}$, which
shows the Painlev\'{e} time coordinates have to exist the same virtual
rotation as the Schwarzschild time coordinates, but this is not seen clearly
in Painlev\'{e} coordinates (\ref{pm}) because on one hand, the Painlev\'{e}
coordinates behave well at horizon and on the other hand, the tranformation
from the Painlev\'{e} coordinates to Kruskal-Szekeres coordinates doesn't have
the same form as (\ref{skt1}) and (\ref{skt2}). Thirdly, there is no
reasonable explanation about the temporal contribution depending on which
direction the horizon is crossed. The momentum is directional but the energy
is not. Otherwise, we also noticed that the rotation $t\rightarrow t-2i\pi M$
will lead to the same result as that $t\rightarrow t+2i\pi M$ for the aim at
connecting these two patches. So whether the temporal contribution from
ingoing and outgoing particles is dependent on the direction has to be
considered carefully. However the temporal contribution indeed exists. Just as
we have pointed out in the last section, the factor of 2 problem is because
the calculation of ingoing amplitude $\left\langle in|out\right\rangle $ is
made only for the spatial change while not included the temporal change. The
Refs. \cite{apgs08,aps08} have presented the temporal change clearly, but we
inclined to think the temporal contribution depending on which direction the
horizon is crossed is due to the interchange between positive energy and
negative energy while the imaginary rotation value is the same in the process.
If so, the ingoing negative energy particles can be regarded as tunneling
inward and the amplitude can be calculated as in Eq. (\ref{net}). Such
explanation is reasonable because the temporal contribution can also be
calculated by treating the ingoing wave amplitude properly, as pointed out in
Refs. \cite{hh76} that the propagator or amplitude at a certain complex value
of $t$ can be obtained by solving the Hamilton-Jabobi equations. Indeed we
find that when calculating the ingoing amplitude one can obtain the temporal
contribution by changing the positive energy into negative energy in
calculation (the normalization included here). So the calculation in our
method is consistent with that in Refs. \cite{apgs08,aps08} based on the third
point of discussion.

Thus, the method suggested in the present paper, which can recover the
standard Hawking temperature by calculating the ingoing and outgoing amplitude
afresh and noting that the ingoing amplitude should be calculated according to
the negative energy particles tunneling inward, is consistent with the other
two methods discussed above. But in our method we use the canonical invariant
tunneling probability and analyze the ingoing and outgoing amplitude
respectively and this can not only related the two other methods but also
supplement or avoid their some deficiencies.

\section{\textbf{Conclusion}}

We have showed that the standard Hawking temperature can be recovered by using
the canonically invariant tunneling probability. In our treatment we find that
the ingoing amplitude should be calculated according to the negative energy
particles tunneling into the black hole and this is because when we change the
spatial direction to calculate the ingoing amplitude, the temporal
transformation have to be considered. In our method, the horizon as two-way
barrier and the Painlev\'{e} coordinates that is proper for discussing the
temperature Hawking radiation as tunneling can be presented clearly. In the
end we also discuss the other two methods and compare them with our method,
which show indeed the radiation temperature $T=\frac{1}{8\pi M}$.

\section{Acknowledgement}

This work is fund by National Natural Science Foundation of China (Grant. No.
10504039 and 10747164).

\end{document}